\documentstyle[prd,tighten,aps]{revtex}
\begin{document}
\draft
\twocolumn[\hsize\textwidth\columnwidth\hsize\csname
@twocolumnfalse\endcsname

\title{Plane waves in quantum gravity: breakdown of the classical
spacetime}
\author{Guillermo A. Mena Marug\'{a}n}
\address{I.M.A.F.F., C.S.I.C., Serrano 121, 28006 Madrid, Spain}
\author{Manuel Montejo}
\address{Universitat de Barcelona, Diagonal 647, 08028
Barcelona, Spain}
\maketitle

\begin{abstract}
Starting with the Hamiltonian formulation for spacetimes with
two commuting spacelike Killing vectors, we construct a
midisuperspace model for linearly polarized plane waves in
vacuum gravity. This model has no constraints and its degrees of
freedom can be interpreted as an infinite and continuous set of
annihilation and creation like variables. We also consider a
simplified version of the model, in which the number of modes is
restricted to a discrete set. In both cases, the quantization is
achieved by introducing a Fock representation. We find
regularized operators to represent the metric and discuss
whether the coherent states of the quantum theory are peaked
around classical spacetimes. It is shown that, although the
expectation value of the metric on Killing orbits coincides with
a classical solution, its relative fluctuations become
significant when one approaches a region where null geodesics
are focused. In that region, the spacetimes described by
coherent states fail to admit an approximate classical
description. This result applies as well to the vacuum of the
theory.
\end{abstract}

\pacs{PACS number(s): 04.60.Ds, 04.30.-w}
\vskip2pc]

\renewcommand{\thesection}{\Roman{section}}
\renewcommand{\theequation}{\arabic{section}.\arabic{equation}}

\section{INTRODUCTION}

The quantization of spacetimes that possess two commuting
spacelike Killing vector fields has received considerable
attention [1-7]. One of the main motivations for the study of
this type of spacetimes is that they generally describe
situations of interest in astrophysics and cosmology \cite{VE}.
Actually, most of the families of spacetimes with two Killing
vectors that have been quantized in the literature can be
interpreted as gravitational waves that propagate either in
Minkowski spacetime or in cosmological universes [1-4]. This is
the case, e.g., of linearly polarized gravitational waves in
cylindrically symmetric spacetimes, which were first analyzed
quantum mechanically by Kucha\v{r} and Allen \cite{KU}. A
consistent quantization of this gravitational system was
achieved by Ashtekar and Pierri \cite{AP}, whilst the most
general model of cylindrical waves in vacuum gravity was
quantized by Korotkin and Samtleben \cite{KS}. Preliminary
discussions on the quantization of gravitational waves in
spacetimes with planar symmetry can be found in Ref. \cite{NE}.
A systematic analysis of purely gravitational plane waves in
quantum geometrodynamics was recently carried out by the authors
\cite{MM}. Finally, the quantization of the Gowdy cosmologies
with the spatial topology of a three-torus was addressed in
Refs. \cite{ME,BE}. These Gowdy spacetimes can be thought of as
inhomogeneous universes with compact sections of constant time
that are filled with gravitational waves \cite{GO}.

Another important motivation for the quantization of spacetimes
with two commuting spacelike Killing vectors comes from their
ability to provide a suitable arena where conceptual issues in
quantum gravity can be advantageously discussed. To date, most
of the gravitational systems that have been quantized to
completion are minisuperspace models \cite{MINI}. Such
gravitational systems are too simple to capture the quantum
field structure of general relativity. In the presence of two
commuting Killing vectors, however, Einstein gravity reduces to
midisuperspace models, namely, gravitational models with an
infinite number of degrees of freedom. It is a common belief
that these models might mimic the complexity that should be
present in a quantum field theory of general relativity.

In this context, the existence of quantum gravitational states
that admit a classical description, or a semiclassical one if
quantum matter is present \cite{WA}, has been recently addressed
\cite{AS,DT}. By considering a model for linearly polarized
gravitational waves with cylindrical symmetry, Ashtekar has
discussed whether it is possible that quantum states are
strongly peaked around classical spacetimes, assuming that the
matter content is given by the expectation value of the
energy-momentum tensor of the matter fields. After a dimensional
reduction, cylindrical waves with linear polarization adopt the
same formulation as axi-symmetric Einstein-Maxwell gravity in
three dimensions \cite{AP,AS}. For the coherent states of the
Maxwell field, it has then been proved that the quantum
fluctuations in the three-dimensional metric are relatively
small only if the coherent state contains neither too many
photons nor photons of high frequency \cite{AS}. In particular,
the three-dimensional metric of the ground state is strongly
peaked around Minkowski spacetime. At least from this
three-dimensional point of view, one can say that the spacetime
foam around the vacuum is smooth. In addition, it has been shown
that, although one can construct states that diminish the
uncertainty in the three-dimensional metric, they induce a loss
of coherence in the Maxwell field \cite{GP}.

The above considerations seem to indicate that, in a certain
sector of quantum gravity, large quantum effects may preclude an
approximate classical description of the spacetime. However,
several points remain obscure in this tentative conclusion. On
the one hand, the discussion in Refs. \cite{AS,GP} has been
carried out from a three-dimensional perspective. Although
axi-symmetric Einstein-Maxwell theory in three dimensions is
equivalent to four-dimensional, cylindrical gravitational waves
with linear polarization, some issues concerning the
interpretation of physical quantities (e.g., the metric) may
depend on the particular approach adopted. On the other hand,
the results obtained could be an artifact of the particular
system studied. In order to discuss the relevance of these
results, one would like to analyze the classical limit of other
midisuperspace models. Actually, large gravitational
fluctuations similar to those described by Ashtekar \cite{AS}
have also been found (again from a three-dimensional
perspective) in a model with toroidal symmetry \cite{BEE}. In
the present work, we will study the quantum behavior of the
four-dimensional metric in another family of spacetimes with two
commuting Killing vectors, namely, the model for linearly
polarized plane waves in vacuum gravity that was discussed in
Ref. \cite{MM}. The analysis of the classical limit for plane
waves is particularly interesting, because these spacetimes show
the remarkable feature of focusing the null cones \cite{PE}. In
the neighborhood of the points where null cones are focused, one
should expect that quantum gravity effects could be especially
important.

The rest of the paper is organized as follows. We first present
our midisuperspace model in Sec. II, where we briefly summarize
the reduction of source-free Einstein gravity carried out in
Ref. \cite{MM} for the case of linearly polarized plane waves.
The degrees of freedom of this model are given by a metric
function $Y$ that, on classical solutions, depends only on one
of the spacetime coordinates, $u$. In Sec. III we show that,
when the field $Y$ corresponds to a flat solution outside a
bounded, fixed interval for the coordinate $u$, it can be
described in terms of an infinite number of discrete modes. The
most general case in which the spacetime is not restricted to be
flat in any region is studied in Sec. IV. The field $Y$ can then
be expanded in a continuous set of modes. The quantization of
these two models, with either discrete or continuous modes, is
discussed in Sec. V. In particular, we introduce regularized
operators that represent the metric functions. In Sec. VI we
define coherent states for the quantum operator associated with
the field $Y$. At least on the orbits of the two spacelike
Killing vectors of the model, the expectation value of the
metric in any coherent state turns out to coincide with a
classical plane-wave solution. We then study the fluctuations in
the metric on such solutions and prove that they become large
when one approaches a region where null cones are focused. As a
consequence, the classical description of the spacetime breaks
down in the vicinity of that region for all coherent states,
including the vacuum. Finally, we discuss our results in Sec.
VII.

\section{LINEARLY POLARIZED PLANE WAVES}
\setcounter{equation}{0}

Purely gravitational plane waves are vacuum solutions to the
Einstein equations that are characterized by possessing as much
symmetry as do plane electromagnetic waves in flat spacetime,
namely, a five dimensional group of motions \cite{BPR}. These
spacetimes are a particular type of plane-fronted gravitational
waves with parallel rays (pp-waves \cite{MAC}) and were first
considered by Baldwin and Jeffrey \cite{BJ}. One can interpret
these waves as describing the gravitational field produced by a
radiating body at great distances \cite{MAC,BJ}.

Although there exists a system of coordinates, called harmonics,
which allows to cover each of the spacetimes for plane waves
with a single chart, the symmetries of the wave are much more
clearly displayed by using group coordinates \cite{VE,MAC}. The
metric can then be written
\begin{equation}\label{Group}
ds^2=-dUdV+h_{ab}(U)dx^adx^b.
\end{equation}
Here, $a,b=1,2$ and the coordinates $x^a$ and $V$ run over the
real line. The coordinate $U$, on the other hand, has a
restricted domain of definition. It is possible to show that, as
$U$ decreases (with a suitable choice of orientation) from any
fixed initial value, a point is reached where the determinant of
the two-metric $h_{ab}$ vanishes \cite{VE} (except in purely
Minkowski spacetime). This point $U_f$ is a coordinate
singularity. As a consequence, group coordinates cannot be
globally employed to describe the whole of the spacetime. The
existence of this coordinate singularity is intimately related
to the focusing effect produced by plane waves \cite{PE}.
Actually, one can prove that null cones are focused on the
hypersurface $U=U_f$.

If the metric function $h_{12}$ vanish, the plane wave is said
to be linearly polarized\cite{MAC}. We will restrict our
discussion to this subfamily of plane waves from now on.

In addition, it can be seen that, modulo a reversal of the
coordinates $U$ and $V$ and a scale transformation of the form
$x^a\rightarrow A x^a$, with $A$ a constant, every gravitational
plane wave presents a region where the determinant of the
two-metric $h_{ab}$ increases with $U$ from zero to the unity
\cite{MM}. In that region, one can perform a change of
coordinates from $U$ to a new coordinate that, in principle,
runs over the whole real axis:
\begin{equation}\label{uU}
u=-\ln{\left(-\frac{1}{2}\ln{[{\rm det} h_{ab}(U)}]\right)}.
\end{equation}
The metric in the considered region adopts then the expression
\cite{MM}
\begin{eqnarray}\label{metric}
ds^2&=&-z_0^{\prime}(u)e^{z_0(u)/2}e^{\Phi(u)}dudV \nonumber
\\ &+&e^{z_0(u)}\left[e^{-y(u)}(dx^1)^2+e^{y(u)}(dx^2)^2\right],
\end{eqnarray}
with
\begin{equation}\label{Phi}
\Phi(u)= \int^u_{u_c}\frac{dr}{2z_0^{\prime}(r)}[y^{\prime}(r)]^2,\;\;\;
\;\;z_0(u)=-e^{-u}.
\end{equation}
Here, $z_0^{\prime}(u)$ and $y^{\prime}(u)$ are the derivatives
of $z_0(u)$ and $y(u)$, respectively, and $u_c$ is a constant.
In terms of $u$, the coordinate singularity that reflects the
focusing effect exerted by the wave has been driven to minus
infinity. Notice also that, from Eqs. (\ref{Group}) and
(\ref{metric}), the explicit expression of the group coordinate
$U$ as a function of $u$ can be obtained, instead of by
inverting relation (\ref{uU}), by integrating the equation
\begin{equation}
dU=z_0^{\prime}(u)e^{z_0(u)/2}e^{\Phi(u)}du.\end{equation}

The metric given above is a solution to the vacuum Einstein
equations for any choice of the arbitrary function $y$. It
describes the most general gravitational plane wave with linear
polarization, except in that it does not represent the whole of
the spacetime that can be covered with harmonic coordinates, but
only a part of it. Apart from neatly displaying the symmetries
of plane waves, the system of coordinates adopted has an
important advantage. After the change $V=2t-u$, metric
(\ref{metric}) can be interpreted by its own as corresponding to
a globally hyperbolic spacetime which possesses two commuting
spacelike Killing vector fields, namely, $\partial_{x^a}$.
Therefore, in order to analyze the quantization of our
gravitational system, one can start with the Hamiltonian
formulation of general relativity for spacetimes with two
commuting Killing vectors. Moreover, for this kind of spacetimes
one can consistently restrict all considerations to the case
that the surface with coordinates $t$ and $u$ is orthogonal to
the group orbits spanned by the Killing vectors, as it happens
for plane waves. This orthogonality condition removes all the
gauge freedom related to diffeomorphisms of the coordinates
$x^a$ \cite{MM}.

To eliminate all the non-physical degrees of freedom and arrive
at a midisuperspace model that describes only linearly polarized
plane waves in source-free gravity, one must introduce
additional gauge-fixing and symmetry conditions, as it is
explained in Ref. \cite{MM}. The reduced system that one attains
in this way is totally free of constraints, has vanishing
reduced Hamiltonian \cite{MM} and its only degrees of freedom
are given by a single field $Y(u)$. In principle, this field may
also depend on the time coordinate $t$; however, since the
dynamical evolution of our reduced model is trivial in the
system of coordinates adopted, $Y$ remains time independent for
all classical solutions. The metric of the model reproduces
expression (\ref{metric}) after the replacements $V=2t-u$ and
\begin{equation}
y(u)=\sqrt{2}e^{-z_0(u)/2}Y(u).
\end{equation}
Hence, the (time independent) classical solutions of our
midisuperspace model are precisely the purely gravitational
plane waves with linear polarization. On the other hand, the
reduced action of the system and the symplectic structure on any
section of constant time are \cite{MM}
\begin{equation}\label{reduced}
S=\int_{t_0}^{t_f}dt\int_{\Sigma}du\, Y^{\prime}
\dot{Y},\;\;\;\;\;\Omega=\int_{\Sigma}du\, {\bf d}Y^{\prime}
\wedge {\bf d}Y.
\end{equation}
Here, $t_0$ and $t_f$ denote the initial and final values of the
time coordinate, the prime stands for the derivative with
respect to the coordinate $u$, and the dot represents the time
derivative (which vanishes only on classical solutions). In
these expressions, surface contributions to the action coming
from sections of constant time have been neglected, and an
overall constant factor has been set equal to the unity by a
suitable choice of the (effective) Newton constant \cite{MM}. In
addition, $\Sigma$ denotes the domain of definition of the
coordinate $u$. In principle, this domain is the real axis.
Nevertheless, it is easy to check that the reduction process
performed in Ref. \cite{MM} can be straightforwardly generalized
to the case that $\Sigma$ is any fixed interval of the real
line.

\section{A MODEL WITH DISCRETE MODES}
\setcounter{equation}{0}

Let us now analyze the case in which the spacetime is flat
outside a fixed region $u\in I_L\equiv [-L,L]$, where $L$ is a
positive constant. To be more precise, we assume that the
function $y(u)$ of the line element (\ref{metric}) takes on
constant values in the region $u>L$ and in $u<-L$. As a
consequence, the same must happen with the metric function
$\Phi(u)$ in Eq. (\ref{Phi}). One can then easily check that,
for $|u|>L$, the spacetime is flat \cite{BPR}. Inside the
interval $I_L$, on the other hand, the function $y(u)$ [and
hence the field $Y(u)$] are left arbitrary. The metrics
considered can be interpreted as describing a kind of sandwich
waves \cite{BPR}. For all practical purposes, one can then
obviate the flat regions $|u|>L$ and replace $\Sigma$ with $I_L$
in the expressions of the reduced action and symplectic
structure (\ref{reduced}).

We will further assume that the plane-wave solutions considered
are sufficiently smooth. In particular, the restriction of
$Y(u)$ to the interval $I_L$ is continuous. The field $Y$ has
then a well-defined limit as $u$ approaches the endpoints of
this interval from its interior (namely, when $u$ tends to
$L^{-}$ and $-L^{+}$). The following argument shows that one can
then restrict the discussion to solutions that satisfy:
\begin{equation}\label{boundary}
\lim_{u\rightarrow L^{-}}Y(u)=\lim_{u\rightarrow -L^{+}}Y(u).
\end{equation}
Suppose that the field $Y(u)$ does not verify this condition. By
adding to $Y$ a suitable function of the form $C e^{z_0(u)/2}$,
where $C$ is a real constant, we can always obtain a new field
$\bar{Y}(u)$ that satisfies Eq. (\ref{boundary}). Moreover, it
is easily checked that the plane-wave metrics obtained with the
fields $Y(u)$ and $\bar{Y}(u)$ differ only by the scale
transformation $x^1\rightarrow A x^1$ and $x^2\rightarrow
x^2/A$, where the constant $A$ equals $e^{-C/\sqrt{2}}$. Since
this scale transformation leaves invariant the domains of
definition of the coordinates $x^a$, which are the real line,
the geometries described by the fields $Y(u)$ and $\bar{Y}(u)$
can be considered equivalent. The boundary condition
(\ref{boundary}) removes then the corresponding overcounting of
equivalent geometries.

Condition (\ref{boundary}) allows us to extend the field $Y(u)$,
restricted to $I_L$, to a periodic function over the entire real
line, its period being equal to $2L$. Since this periodic
function coincides with our field in the whole interval of
interest, $I_L$, we will denote it also by $Y(u)$, to keep the
notation as simple as possible. The periodicity (and smoothness)
of $Y(u)$ guarantees that it can be expanded in the following
Fourier series:
\begin{equation}\label{Fouriers}
Y(u)\!=a_0+\!\sum_{n=1}^{\infty}\frac{1}{2\sqrt{\pi n}}\left(
a_n e^{-in\pi u/L}+\!a^{\star}_n e^{in\pi u/L}\right)\!.
\end{equation}
In order for $Y(u)$ to be real, $a^{\star}_n$ must be the
complex conjugate of $a_n$, and $a_0$ must be real. The Fourier
coefficients in this expression may, in principle, depend on the
coordinate $t$, although they become time independent on
classical solutions (recall that the reduced Hamiltonian
vanishes).

Employing condition (\ref{boundary}) and Eq. (\ref{reduced}), it
is not difficult to show that the Fourier coefficient $a_0$
disappears from the expressions of the reduced action and
symplectic structure of the model. Therefore, the zero mode
decouples from the other degrees of freedom of the system. In
particular, this implies that we can consistently restrict all
considerations to the case in which $a_0$ takes a specific
value, independent of the coordinate $u$. We can take advantage
of this fact to demand that the field $Y$ vanishes at a certain
point $u_0$ of the interval $I_L$. In other words, we can set
$a_0=-Y(u_0)$. Note that, with this choice, we are selecting the
value of $Y$ at $u_0$ as the reference value with respect to
which the field $Y$ is going to be measured. The Fourier series
(\ref{Fouriers}) becomes then
\begin{eqnarray}\label{bilocal}
Y(u|u_0)&=&\sum_{n=1}^{\infty}\frac{1}{2\sqrt{\pi n}}\left[ a_n
\left(e^{-in\pi u/L}-e^{-in\pi u_0/L}\right)\right.\nonumber \\
&+& \left. a^{\star}_n \left(e^{in\pi u/L}-e^{in\pi
u_0/L}\right)\right],
\end{eqnarray}
which can be interpreted as a bilocal field (i.e., the
difference between the values of a field at two points). In the
above equation, we have made explicit the dependence on the
point $u_0$. On the other hand, substituting the above series in
the reduced symplectic structure and recalling that the zero
mode decouples from the system, we obtain $\Omega=-i\sum_{n=1}
^{\infty}{\bf d}a_n\wedge {\bf d}a^{\star}_n$. Hence, the only
non-vanishing Poisson brackets between the Fourier coefficients
are $\{a_n,a^{\star}_m\}=-i\delta_n^m$, where $\delta$ denotes
the Kronecker delta. Since the coefficients $a_n$ ($n\geq 1$)
are generally complex and conjugate to $a^{\star}_n$, the
degrees of freedom of our model can then be interpreted as an
infinite and discrete set of annihilation and creation like
variables.

The quantization of the model can be achieved by introducing a
Fock representation, as we discuss in Sec. V. In order to avoid
ultraviolet divergences in that quantization, the operator that
represents the field $Y$ needs to be regularized. From a
physical point of view, such a regularization can be justified
as follows. In any real measurement of the (bilocal) field
$Y(u|u_0)$, the positions of the points $u$ and $u_0$ will not
be determined with total accuracy. One rather expects that the
measurement will be performed over a small neighborhood of each
of these points; the result will be an average of the form
\begin{equation}\label{Ymeas}
\!Y_R(u|u_0)\!=\!\!\int_{I\!\!\!\,R}\!\!\! d\bar{u}
g(u-\bar{u})\!\int_{I\!\!\!\,R}\!\!\!d\bar{u}_0 g(u_0-\bar{u}_0)
Y(\bar{u}|\bar{u_0}),
\end{equation}
where $g(u)$ is a smooth [$C^{\infty}(I\!\!\!\,R)$] test
function of small compact support and unit integral,
$\int_{I\!\!\!\,R} du g(u)=1$.

Some comments are in order, concerning this formula. Since the
original field $Y$ (which is the physically relevant one)
coincides with its periodic extension only in the interval
$I_L$, the same applies to the bilocal field $Y(u|u_0)$ with
respect to its two arguments. As a consequence, the bilocal
field in expression (\ref{Ymeas}) can be substituted by the
Fourier series (\ref{bilocal}) only if the integrals in that
expression can be restricted to the interval $I_L$. Let then
$I_{\epsilon}\equiv [-\epsilon,\epsilon]$ be the support of $g$,
where, according to our previous discussion, $\epsilon\ll L$.
This last condition guarantees that our measurements are good
enough as to differentiate a huge number of regions in $I_L$. It
is then easy to check that, as far as the points $u$ and $u_0$
belong to the interval $I_{L-\epsilon}\equiv
[-L+\epsilon,L-\epsilon]$, the integrals in Eq. (\ref{Ymeas}) do
not receive contributions from regions outside $I_L$.

Substituting then the Fourier expansion of $Y(u|u_0)$ and
recalling the properties of the function $g$, one obtains
\begin{equation}\label{Yg}
Y_R(u|u_0)=\frac{1}{\sqrt{2}}\sum_{n=1}^{\infty}
\left[f^{\star}_n(u|u_0)a_n+f_n(u|u_0) a^{\star}_n\right],
\end{equation}
\begin{equation}\label{fn}
f_n(u|u_0)=\frac{e^{in\pi u/L}-e^{in\pi u_0/L}}{\sqrt{n}}
\tilde{g}\left(\frac{n\pi}{L}\right),
\end{equation}
$f^{\star}_n$ being the complex conjugate of $f_n$ and
$\tilde{g}$ denoting the Fourier transform of $g$\cite{RS}. For
points $u$ or $u_0$ whose absolute value lies between
$L-\epsilon$ and $L$, Eqs. (\ref{Yg}) and (\ref{fn}) must be
modified. In the following, however, we will only consider the
case $u,u_0\in I_{L-\epsilon}$, i.e., we will assume that all
measurements of the field are made in the interior of the
interval $I_L$ and sufficiently far from its endpoints.

On the other hand, since $g(u)$ is a smooth function of compact
support, its Fourier transform belongs to ${\cal
S}(I\!\!\!\,R)$, the Schwartz space of smooth test functions of
rapid decrease \cite{RS}. Using this property, it is not
difficult to prove that the sequence $f\equiv\{f_n; n\geq 1\}$
is square summable, $f\in l^2$. This fact will be employed in
Sec. V to introduce a well-defined operator that represents the
measured field $Y_R(u|u_0)$. Let us finally comment that, if we
had not performed an average over the positions of $u$ and
$u_0$, expressions (\ref{Yg}) and (\ref{fn}) would still have
been valid for the bilocal field $Y(u|u_0)$, but with the
function $\tilde{g}$ replaced with the unit function divided by
$\sqrt{2\pi}$. In that case, the resulting sequence $f$ would
have not belonged to the Hilbert space $l^2$, owing to the
divergent contribution of the high-frequency modes ($n\gg 1$).

\section{CONTINUOUS MODE EXPANSION}
\setcounter{equation}{0}

We now return to the general case in which the field $Y(u)$
describes an arbitrary plane wave with linear polarization. As a
boundary condition, we will demand the field to be of order
unity at plus and minus infinity (we will say that a function
$f(u)$ is of order $g(u)$ at $u_0$, and write $f(u)=O[g(u)]$ as
$u\rightarrow u_0$, if the limit of $f/g$ exists at $u_0$). In
this situation, similar arguments to those presented for the
model with discrete modes show that, in order to avoid
overcounting of equivalent geometries, one can restrict all
considerations to fields that satisfy
\begin{equation}\label{Boundary}
\lim_{u\rightarrow \infty} Y(u)=\lim_{u\rightarrow -\infty}
Y(u).
\end{equation}

Let us next extract from $Y$ its value at infinity. This value
can be regarded as a function of the coordinate $t$ that, like
the field $Y$, remains constant on classical solutions.
Employing condition (\ref{Boundary}), one can check that this
function has vanishing contribution to the reduced action and
symplectic structure of the system, as it happened with the
Fourier coefficient $a_0$ in the model of Sec. III. Thanks to
this decoupling, one can consistently analyze any sector of the
space of solutions where the considered function takes a
specific value. We will benefit from this fact and limit our
discussion to fields $Y(u|u_0)$ that vanish at a given point,
$u_0$, as we did in the previous section. This restriction can
again be interpreted as the choice of a reference value for the
field $Y$.

We will finally assume that, apart from its constant value at
infinity, the field $Y$ can be expressed as a Fourier transform.
This assumption is the analogue of the expansion of $Y$ in a
Fourier series employed in Sec. III. Recalling the restriction
to fields that vanish at $u_0$, we can then write
\begin{eqnarray}\label{Bilocal}
Y(u|u_0)&=&\frac{1}{2\sqrt{\pi}}\int_{0}^{\infty}
\frac{dk}{\sqrt{k}}
\left[ a(k) \left(e^{-iku}-e^{-iku_0}\right)\right.\nonumber \\
&+&\left.a^{\star}(k)\left(e^{iku}-e^{iku_0}\right)\right].
\end{eqnarray}
Since $Y(u|u_0)$ is real, the functions $a(k)$ and
$a^{\star}(k)$ must be complex conjugate to each other. On the
other hand, the boundary condition (\ref{Boundary}) can now be
translated into conditions on the function $a(k)$. Let us define
\begin{equation}
b(k)\!=\theta(k) a(k)+\theta(-k) a^{\star}(|k|),\;\;\;\;\;
c(k)\!=\!\frac{b(k)}{\sqrt{2|k|}},
\end{equation}
where $\theta$ is the Heaviside step function. It is then easy
to check that $Y(u|u_0)$ equals $\tilde{c}(u)-\tilde{c}(u_0)$,
where $\tilde{c}$ is the Fourier transform of $c$. Therefore, it
suffices that $c$ is absolutely integrable [$c\in
L^1(I\!\!\!\,R)$] to guarantee that condition (\ref{Boundary})
is fulfilled, because in that case $\tilde{c}$ is a continuous
function that vanishes at infinity \cite{RS}. In particular, one
can prove that $c$ belongs to $L^1(I\!\!\!\,R)$ provided that
$a(k)$ is square integrable over the positive real axis [$a\in
L^2(I\!\!\!\,R^+)$] and there exist constants $\alpha>1/2$ and
$\beta<1/2$ such that
\[
a(k)=O[k^{-\alpha}] \;\;{\rm as}\;\; k\rightarrow\infty,\;\;\;
a(k)=O[k^{-\beta}] \;\; {\rm as}\;\; k\rightarrow 0.
\]
All these requirements are satisfied, e.g., if the function
$b(k)$, defined above, belongs to the Schwartz space ${\cal
S}(I\!\!\!\,R)$.

We can regard the complex conjugate functions $a(k)$ and
$a^{\star}(k)$ as the degrees of freedom of our model. In
principle, these functions might depend on the time coordinate
$t$; however, since the field $Y$ remains constant on classical
solutions, they are classically time independent. In addition,
replacing expression (\ref{Bilocal}) in the symplectic structure
[and recalling condition (\ref{Boundary})], one obtains
\begin{equation}
\Omega=-i\int_0^{\infty} dk\, {\bf d}a(k)\wedge{\bf
d}a^{\star}(k),
\end{equation}
so that the only non-vanishing Poisson brackets between our
variables are $\{a(k),a^{\star}(\bar{k})\}=-i\delta(k-\bar{k})$.
The degrees of freedom for linearly polarized plane waves can
thus be interpreted as a continuous set of annihilation and
creation like variables, whose quantization can be carried out
by introducing a Fock representation.

As we argued in Sec. III, any physical measurement of the
(bilocal) field $Y(u|u_0)$ would imply an average of the form
(\ref{Ymeas}) over the positions of $u$ and $u_0$. Employing Eq.
(\ref{Bilocal}), this average leads to
\begin{equation}
Y_R(u|u_0)=\!\int_0^{\infty}\frac{dk}{\sqrt{2}}\left[
f^{\star}(k,u|u_0)a(k)+f(k,u|u_0)a^{\star}(k)\right],
\end{equation}
where, for $k>0$,
\begin{equation}\label{Fn}
f(k,u|u_0)=\frac{e^{iku}-e^{iku_0}}{\sqrt{k}}\tilde{g}(k).
\end{equation}
It is worth noting that the above expressions can be obtained
from Eqs. (\ref{Yg}) and (\ref{fn}) in the continuum limit (when
$L$ and $n$ tend to infinity, keeping $k=n\pi /L$ finite), a
fact that supports our conclusions. Notice also that, in the
limit $L\rightarrow\infty$, the region where the spacetime is
flat for the model of Sec. III is driven to infinity, whereas
the interval $I_{L-\epsilon}$ [namely, the region where formulas
(\ref{Yg}) and (\ref{fn}) are valid] becomes the real line. On
the other hand, since $g$ is a smooth test function of compact
support, its Fourier transform $\tilde{g}$ belongs to the
Schwartz space ${\cal S}(I\!\!\!\,R)$. It then follows that
$f(k,u|u_0)$ is a square integrable function over the positive
real line for all finite values of $u$ and $u_0$. This property
will be used in the next section to attain a well-defined
operator that represents the field $Y_R$ quantum mechanically.

To close this section, let us comment that, under the change of
coordinates $u=x+t$, the field $Y(u|u_0)$ can actually be
interpreted as a bilocal field constructed from the left-mover
part of a massless scalar field in two dimensions (those
corresponding to the coordinates $x$ and $t$) \cite{INFRA}. It
is well known that such a scalar field presents infrared and
ultraviolet divergences \cite{INFRA}. In our model, however, the
infrared divergences have been eliminated by considering bilocal
fields, whilst the average over positions has taken care of the
ultraviolet divergences. This explains why the function
$f(k,u|u_0)$ is square integrable: the convergence near $k=0$
has been achieved by subtracting contributions from the point
$u_0$, whereas the convergence at infinity is ensured by the
decay of $\tilde{g}$.

\section{QUANTUM THEORY}
\setcounter{equation}{0}

We have argued that, for the two models considered in this work,
a measurement of the field $Y$ (assumed to vanish at the point
$u_0$) should lead to a result of the form:
\begin{equation}\label{field}
\!Y_R(u|u_0)\!=\!\frac{1}{\sqrt{2}}[<\!a,f(u|u_0)\!>\!+\!
<\!a^{\star},f^{\star}(u|u_0)\!>].
\end{equation}
Here, $f$ and $f^{\star}$ are two vectors of a complex Hilbert
space, ${\cal H}$, which are complex conjugate to each other and
depend on the points $u$ and $u_0$. In addition, $<,>$ denotes
the inner product on ${\cal H}$, defined as an antilinear
mapping of its second argument. For the discrete model analyzed
in Sec. III, ${\cal H}$ is the Hilbert space $l^2$ of square
summable sequences, $f(u|u_0)$ denotes the sequence
$\{f_n(u|u_0);\;n\geq 1\}$ given in Eq. (\ref{fn}), and $a$ and
$a^{\star}$ stand for the discrete set of annihilation and
creation like variables $\{a_n;\;n\geq 1\}$ and
$\{a^{\star}_n;\; n\geq 1\}$, respectively. For the general case
of linearly polarized plane waves, on the other hand, ${\cal H}$
is the Hilbert space $L^2(I\!\!\!\,R^+)$, $f(u|u_0)$ is the
function of $k\in I\!\!\!\,R^+$ defined in Eq. (\ref{Fn}), and
$a$ and $a^{\star}$ denote the continuous set of annihilation
and creation like variables $a(k)$ and $a^{\star}(k)$.

The quantization of the (bilocal) field $Y_R(u|u_0)$ can then be
achieved by constructing a Fock representation of the
annihilation and creation like variables. We will call ${\cal
H}^{(n)}$ the tensor product $\otimes_{m=1}^n{\cal H}$. Let also
$S_n$ be the symmetrization operator defined on ${\cal H}^{(n)}$
by its action on states of the form
$\phi^{(n)}\equiv\phi_1\otimes
\cdots\otimes\phi_n$:
\begin{equation}\label{otimes}
S_n\phi^{(n)}=\sum_{\sigma}\frac{1}{n!}\phi_{\sigma(1)}\otimes
\cdots\otimes\phi_{\sigma(n)}\equiv\phi_s^{(n)},
\end{equation}
where the summation is over all permutations $\sigma$. The
symmetric Fock space over ${\cal H}$ is ${\cal F}_s({\cal
H})=\oplus_{n=0}^{\infty}{\cal H}_s^{(n)}$, where ${\cal
H}_s^{(n)}=S_n{\cal H}^{(n)}$ is called the $n$-th particle
subspace \cite{RS}. Let finally ${\cal F}_0\subset {\cal
F}_s({\cal H})$ be the dense subspace of finite particle
vectors, i.e., the subspace of vectors
$\phi_s\equiv\{\phi_s^{(n)};\;n\geq 0\}$ such that
$\phi_s^{(n)}$ vanishes for all but a finite set of indices $n$.
We can then define an annihilation operator $\hat{a}(f)$ on
${\cal F}_s({\cal H})$, with domain ${\cal F}_0$, via its action
on vectors of the type (\ref{otimes}):
\[
\hat{a}(f)\phi_s^{(n)}=\sum_{\sigma}\frac{\sqrt{n}}{n!}
<\phi_{\sigma(1)},f>\phi_{\sigma(2)}\otimes\cdots\otimes
\phi_{\sigma{(n)}}
\]
with $n\geq 1$, whereas $\hat{a}(f)$ vanishes on ${\cal
H}_s^{(0)}$ \cite{RS}. This operator represents the classical
quantity $<a,f>$ and is closable for any $f\in{\cal H}$. Its
adjoint is the creation operator $\hat{a}^{\star}(f)$, which
represents the variable $<a^{\star},f^{\star}>$. Its restriction
to the subspace of finite particle vectors can be obtained from
the formula \cite{RS}:
\begin{equation}
\hat{a}^{\star}(f)\phi_s^{(n)}=\sqrt{n+1}\,S_{n+1}
\left(f\otimes \phi_s^{(n)}\right).
\end{equation}
Notice that, from our definitions, $\hat{a}^{\star}(f)$ is
linear in $f$, whereas $\hat{a}(f)$ is antilinear.

It is possible to understand each vector $\phi_s^{(n)}$ in
${\cal H}_s^{(n)}$ as a quantum state $|\phi_s^{(n)}>$ obtained
from a vacuum $|0>$ by the action of $n$ creation operators. For
vectors given by formula (\ref{otimes}), e.g., one has:
\begin{equation}
|\phi_s^{(n)}
>=\frac{1}{\sqrt{n!}}\hat{a}^{\star}(\phi_{1})
\cdots\hat{a}^{\star}(\phi_{n})|0>.
\end{equation}
The constant overall factor is a normalization constant
introduced for convenience. The vacuum $|0>$ is characterized as
the only state that is destroyed by all annihilation operators
and has unit norm. It corresponds to the vector
$\phi_s^{(0)}=1$.

For any vector $f$ in ${\cal H}$, let us next introduce the
Segal field operator $\hat{Y}_R$ on ${\cal F}_0$ \cite{RS},
defined by
\begin{equation}
\hat{Y}_R[f]=\frac{1}{\sqrt{2}}\left(\hat{a}(f)+
\hat{a}^{\star}(f)\right).
\end{equation}
In particular, the operator $\hat{Y}_R[f(u|u_0)]$ represents the
bilocal field (\ref{field}). The Segal field operator on ${\cal
F}_0$ satisfies the commutation relations
\begin{equation}\label{commutator}
\left[\hat{Y}_R[f],\hat{Y}_R[g]\right]= i {\rm Im}<g,f>,
\end{equation}
where ${\rm Im}$ denotes the imaginary part and we have set
$\hbar=1$. This operator has a self-adjoint closure \cite{RS},
which we will also call $\hat{Y}_R$. The spectral theorem
\cite{RSI} ensures then that $\exp{(A\hat{Y}_R[f])}$ is a
well-defined, self-adjoint, and positive operator for all
vectors $f\in {\cal H}$, where $A$ is any real $c$-number.

Therefore, after replacing the field $Y(u)$ with its bilocal
version $Y_R(u|u_0)$, we can represent the diagonal components
$h_{aa}(u)$ ($a=1$ or 2) for the plane-wave metrics
(\ref{metric}) by the regularized, positive operators:
\begin{eqnarray}\label{Ometric}
\hat{h}_{aa}^R(u|u_0)&=&\exp{z_0(u)}
\exp{\left\{-e^{-z_0(u)}||f(u|u_0)||^2/2\right\}}\nonumber\\
&\times&\exp{\left\{(-1)^a\sqrt{2}e^{-z_0(u)/2}\hat{Y}_R[f(u|u_0)]
\right\}}.
\end{eqnarray}
Here, $||f||$ is the norm of $f\in{\cal H}$ and we have
displayed the dependence on the points $u$ and $u_0$. We recall
that, for the discrete model discussed in Sec. III, the points
$u$ and $u_0$ have been restricted to lie in the interval
$I_{L-\epsilon}$, whilst for the general case of linearly
polarized waves analyzed in Sec. IV, $u$ and $u_0$ can take any
real value. The second factor on the r.h.s. of this expression
can be understood as follows. We choose to normal order the
exponential of $\hat{Y}_R[f(u|u_0)]$, so that the vacuum
expectation value of the operators (\ref{Ometric}) reproduces
the classical value of the two-metric $h_{ab}$ when the field
$Y$ vanishes, i.e., in flat spacetime. Employing then that (on
${\cal F}_0$)
\begin{equation}\label{acommuta}
[\hat{a}(f),\hat{a}^{\star}(g)]= <g,f>
\end{equation}
and the Campbell-Baker-Hausdorff (CBH from now on) formula
$e^{\hat{b}}e^{\hat{c}}=e^{[\hat{b},\hat{c}]/2}e^{(\hat{b}+
\hat{c})}$,
valid for operators $\hat{b}$ and $\hat{c}$ whose commutator is
a $c$-number \cite{MW}, we finally arrive at the above
expression for $\hat{h}^R_{aa}$.

We will not analyze in detail the introduction of an operator
that represents the other independent component of the metric
(\ref{metric}), which (after the change of coordinates $V=2t-u$)
can be written in the form
$h_{uu}=z_0^{\prime}(u)e^{z_0(u)/2}e^{\Phi(u)}$, the function
$\Phi(u)$ being given by Eq. (\ref{Phi}). As we will see in the
next section, the discussion about the existence of large
quantum effects that invalidate the classical description of the
metric can be carried out without relying on a particular
definition for this operator. Let us simply comment that
\begin{equation}
\!\hat{\Phi}_R(u|u_0)\!=\!\int_{u_c}^u\!\frac{dr}{z_0^{\prime}(r)}
:\!\left[\!\left(e^{-z_0(r)/2}\hat{Y}_R[f(r|u_0)]\right)^{\prime}
\right]^2\!\!\!\!:
\end{equation}
can be proved to be a densely defined operator on the symmetric
Fock space ${\cal F}_s({\cal H})$. Therefore, if
$\hat{\Phi}_R(u|u_0)$ is essentially self-adjoint, its
exponential [multiplied by the function
$z_0^{\prime}(u)e^{z_0(u)/2}$] provides a well-defined, positive
operator that could be interpreted as a quantum counterpart of
$h_{uu}$. Notice that $\hat{\Phi}_R(u|u_0)$ has been obtained
from the expression of $\Phi(u)$ by replacing the classical
field $Y(u)$ with the Segal field operator $\hat{Y}_R[f(u|u_0)]$
and taking normal ordering.

\section{METRIC FLUCTUATIONS}
\setcounter{equation}{0}

In this section, we will discuss whether there exist quantum
states in our model that admit a classical description of the
spacetime. It seems natural to consider as candidates the
coherent states \cite{MW} of the basic field operator
$\hat{Y}_R[f]$, since all these states are strongly peaked
around classical field solutions. Actually, most of the analysis
of large quantum gravity effects presented in the literature has
been carried out by investigating the geometry fluctuations on
states of this type \cite{AP,AS,BEE}. In the models that have
been considered, however, the geometry studied was obtained by a
Killing reduction to three-dimensions. Here, we will study the
quantum behavior of linearly polarized plane waves from a purely
four-dimensional point of view. We are particularly interested
in discussing the metric fluctuations on the vacuum, which is
the coherent state that should represent flat spacetime.

Given any vector $c$ in the Hilbert space ${\cal H}$ (where
${\cal H}$ is $l^2$ for the model of Sec. III and
$L^2(I\!\!\!\,R^+)$ in the general case of linearly polarized
waves), we define the coherent state $|c>$ by
\begin{equation}
|c>=\exp{(-||c||^2/2)}\,\exp{\hat{a}^{\star}(c)}\, |0>.
\end{equation}
The overall numerical factor guarantees that $|c>$ has unit
norm. Using formula (\ref{acommuta}) and recalling that the
vacuum is destroyed by all annihilation operators, it is not
difficult to see that, for any $f\in {\cal H}$, the expectation
value of the Segal field on any coherent state is
\begin{equation}\label{fieldmv}
<\hat{Y}_R[f]>_c=
\frac{1}{\sqrt{2}}\left(<c,f>+<f,c>\right).
\end{equation}
Regardless of the value of $f$, this expectation value coincides
with the classical field obtained by replacing the annihilation
variables $a$ with the vector $c\in {\cal H}$ (and $a^{\star}$
with its complex conjugate, $c^{\star}$). On the other hand, for
every fixed vector $f\in{\cal H}$, let us call $\hat{Y}_R[if]$
the canonical momentum of the field $\hat{Y}_R[f]$. Notice that,
from Eq. (\ref{commutator}), the commutator of these two
operators is the imaginary $c$-number $i||f||^2$. One can then
check that all coherent states have the same uncertainty in the
field $\hat{Y}_R[f]$ and in its canonical momentum. Furthermore,
for all vectors $f\in {\cal H}$, the product of these
uncertainties attains its minimum value \cite{JA}, namely,
$||f||^2/2$.

Let us now analyze the quantum fluctuations in the metric. We
will restrict our considerations to the operators
(\ref{Ometric}), which describe the metric on the Killing orbits
of our midisuperspace model. Our conclusions will be independent
of the quantum behavior of $h_{uu}$, i.e., the other non-trivial
component of the metric for plane waves with linear
polarization. By defining
\begin{equation}\label{Ametric}
F(u)=e^{z_0(u)},\;\;\;\;\; A_a(u)=(-1)^a\sqrt{2}F(u)^{-1/2}
\end{equation}
(with $a=1,2$) and suppressing from our notation the explicit
dependence on the points $u$ and $u_0$, we can write Eq.
(\ref{Ometric}) in the compact form
\begin{equation}
\hat{h}^R_{aa}=F\,:\exp{(A_a\hat{Y}_R[f])}:\,,
\end{equation}
where the dots denote normal ordering. Employing the CBH formula
and Eq. (\ref{acommuta}), one finds that
\begin{equation}\label{expcoh}
<\!F\!:\exp{(A_a\hat{Y}_R[f])}\!:>_c=\!F \exp{(\sqrt{2}A_a
{\rm Re}<\!c,f\!>)}.
\end{equation}
Here, we have used that $A_a$ is a real $c$-number, and ${\rm
Re}$ stands for the real part. We therefore see that the
expectation value of $\hat{h}^R_{aa}$ on a coherent state
reproduces the classical value obtained from the field solution
(\ref{fieldmv}). In particular, for the vacuum state one gets
the flat spacetime value $<\hat{h}^R_{aa}>_0=F\delta_a^a$.

From Eq. (\ref{expcoh}), it is straightforward to conclude that
\begin{equation}\label{uncertmet}
\left(\frac{\Delta_c \hat{h}^R_{aa}}{<\hat{h}^R_{aa}>_c}\right)^2
=\exp{\left(\frac{A_a^2||f||^2}{2}\right)}-1.
\end{equation}
For any observable $\hat{b}$, the symbol $\Delta_c \hat{b}$
denotes the uncertainty of the state $|c>$, namely, the square
root of the difference between $<\hat{b}^2>_c$ and
$(<\hat{b}>_c)^2$. It is worth remarking that, from the above
expression, the relative fluctuations in the two-metric
$\hat{h}^R_{aa}$ turn out to be independent of the particular
coherent state considered. Note also that there will generally
exist large fluctuations in the geometry whenever the
expectation value of the metric $\hat{h}^R_{aa}$ is large.
However, approximating this metric by the classical value
(\ref{expcoh}) will still be acceptable provided that the
relative fluctuations in $\hat{h}^R_{aa}$ are small. According
to Eqs. (\ref{Ametric}) and (\ref{uncertmet}), this will be the
case if and only if
\begin{equation}\label{class}
e^{-z_0(u)}||f(u|u_0)||^2\ll 1.
\end{equation}

Before continuing our discussion, we would like to comment on
the implications that the uncertainty principle may have for
operators of the form $\hat{h}^R_{aa}$, which are given by the
normal ordered exponential of a Segal field multiplied by a
positive $c$-number. Remember that $\hat{h}^R_{aa}$ denotes a
different operator for each pair of points $u$ and $u_0$. Let
then $\hat{b}=F:e^{\hat{Y}_R[f]}:$ and
$\hat{c}=G:e^{\hat{Y}_R[g]}:$ be two such operators. We know
that the uncertainty product in $\hat{b}$ and $\hat{c}$ is
always greater than half the norm of the expectation value of
$[\hat{b},\hat{c}]$ \cite{JA}. In our case, the commutator
$[\hat{b},\hat{c}]$ is a quantum operator proportional to
$FG:e^{\hat{Y}_R[f+g]}:$$\,$ Hence, its expectation value will
depend on the state analyzed. As a consequence, it is not clear
which is the minimum value allowed for the product of
uncertainties in $\hat{b}$ and $\hat{c}$. In this situation, it
seems more natural to consider, e.g., the uncertainty product
divided by the expectation value of $FG:e^{\hat{Y}_R[f+g]}:$$\,$
This quantity turns out to be bounded by the $c$-number
$|e^{<g,f>/2}-e^{<f,g>)/2}|/2$ and, on coherent states,
coincides with the product of the relative fluctuations in
$\hat{b}$ and $\hat{c}$, as can be checked by employing Eq.
(\ref{expcoh}). Using techniques explained in Ref. \cite{JA},
one can then prove that the considered quantity is not minimized
by the coherent states for generic vectors $f,g\in {\cal H}$.

Let us now return to the analysis of the metric fluctuations.
For the whole family of coherent states, we have seen that the
classical description of the spacetime may be acceptable only if
condition (\ref{class}) is satisfied at all points $u$. Recall
that $u_0$ is a conveniently chosen point where the diagonal
components of the metric in the $x^a$-directions ($a=1,2$) are
set equal to $e^{z_0(u_0)}$ by convention, both classical and
quantum mechanically. For the discrete model of Sec. III, the
vector $f(u|u_0)$ denotes the sequence (\ref{fn}) in $l^2$, and
for the model of Sec. IV, $f(u|u_0)$ is the function (\ref{Fn}),
which belongs to $L^2(I\!\!\!\,R^+)$. In general, inequality
(\ref{class}) will not be satisfied if the norm of the vector
$f(u|u_0)$ or the function $e^{-z_0(u)}$ becomes considerably
large at a certain point $u$. Since the form of $f(u|u_0)$ (and
hence its norm) may depend on the regularization adopted in the
quantization, we will concentrate our discussion on the
possibility that $e^{-z_0(u)}$ is significantly large, a
possibility that is insensitive to the ambiguities found in the
construction of the quantum theory. Given the definition of
$z_0$ in Eq. (\ref{Phi}), the function $e^{-z_0(u)}$ is the
double exponential of $-u$. This exponential becomes unbounded
from above when $u$ approaches $-\infty$, which is precisely the
region where null cones are focused by the plane wave, as we
commented in Sec. II. We hence expect the quantum spacetime to
radically differ from its classical approximation close to this
focusing region.

In order to prove this statement, we only have to show that the
norm of $f(u|u_0)$ remains strictly positive when $u\rightarrow
-\infty$. It then follows that $||f(u|u_0)||^{-2}$ is bounded as
a function of $u$ away from $u_0$ because, for every possible
choice of the smearing function $g$ employed in our
regularization, the vector $f(u|u_0)$ is normalizable and
different from zero at all finite points $u\neq u_0$. As a
consequence, condition (\ref{class}) will not be satisfied when
$u$ becomes large and negative \cite{note}.

Let us first analyze the model with discrete modes. We set
$u_0=0$ and $u=-L/3$ for convenience. In particular, if $L\gg
\epsilon$, the points $u$ and $u_0$ belong to the interval
$[-L+\epsilon,L-\epsilon]$ (to which we restricted our
considerations in Sec. III). Using expression (\ref{fn}), it is
not difficult to check that
$||f(-L/3\,|0)||>|\tilde{g}(\pi/L)|$. Since $\tilde{g}$ is the
Fourier transform of a smooth function with unit integral, we
have that $\tilde{g}(0)=1/\sqrt{2\pi}$. In the limit
$L\rightarrow\infty$, when the point $u=-L/3$ approaches the
region where null geodesics are focused, the norm of
$f(-L/3\,|0)$ is then bounded from below by $1/\sqrt{2\pi}$.
Therefore, in the case with discrete modes studied in Sec. III
and assuming that the value of $L$ is sufficiently large, the
relative fluctuations in the metric turn out to be huge when
$u\ll 0$, regardless of the specific form taken by the function
$\tilde{g}$. Thus, the classical spacetime breaks down for all
coherent states in the region $0\gg u>-L+\epsilon$.

In the general model of plane waves with linear polarization,
one can also show that, in the limit $u\rightarrow -\infty$, the
norm of the function $f(u|u_0)$ is bounded from below by a
strictly positive number \cite{note,nota} for every possible
choice of the function $g$ used in the regularization. Hence, we
conclude again that the relative uncertainty in the metric
$\hat{h}^R_{aa}$ is large close to the hypersurface where null
cones are focused, i.e, for large and negative values of the
coordinate $u$. In that region at least, the classical
description of the spacetime, as a plane wave with
$x^a$-components of the classical metric given by Eq.
(\ref{expcoh}), is not acceptable for any of the coherent
states. Notice that this breakdown of the classical spacetime
when one approaches the focusing region is regulator
independent, because it occurs for all of the admissible choices
of smearing function $g$. Finally, let us emphasize that our
results apply as well to the vacuum, which is the quantum state
that should describe flat spacetime. So, quantum gravitational
effects around the vacuum cannot be neglected in the vicinity of
the region where null geodesics are focused by linearly
polarized plane waves.

\section{SUMMARY AND DISCUSSION}
\setcounter{equation}{0}

Linearly polarized plane waves in source-free gravity have been
described by a midisuperspace model that is free of constraints
and whose reduced Hamiltonian vanishes. This model was obtained
in Ref. \cite{MM}, starting with the Hamiltonian formulation of
general relativity for spacetimes with two commuting spacelike
Killing vectors and introducing gauge-fixing and symmetry
conditions. The degrees of freedom of the model are given by a
field $Y$ that (on classical solutions) depends only on one of
the spacetime coordinates, namely, the coordinate $u$. We have
shown that one can consistently restrict all considerations to
fields that vanish at a given point $u_0$. Furthermore, after a
Fourier expansion, this field can be expressed in terms of an
infinite collection of annihilation and creation like variables.
In the general case of linearly polarized waves, these variables
form a continuous set, but we have also considered a simplified
version of the model in which the field $Y$ describes flat
spacetime outside a fixed region. If one is only interested in
studying that region, the Fourier expansion of the field can be
restricted to a discrete set of modes. Finally, we have argued
that the position of the points $u$ and $u_0$ cannot be
determined with total accuracy in any physical measurement. As a
consequence, the field measured, $Y_R(u|u_0)$, would result from
an average over some small neighborhood of those points.

In order to quantize the system, we have introduced a Fock
representation and described the field $Y_R(u|u_0)$ by a Segal
field operator. This Segal quantization has been carried out
over the Hilbert space of square summable sequences for the
model with discrete modes, and over the space of square
integrable functions on the positive real line for the general
case of plane waves with linear polarization. In this way, we
have been able to construct regularized operators that represent
the spacetime metric for plane waves. At least for the
two-metric on Killing orbits (i.e., the sections of constant
coordinates $t$ and $u$), we have proved that these regularized
operators are self-adjoint and positive. We have then
concentrated our discussion on the analysis of the coherent
states of the quantum theory. These states are characterized by
minimizing the uncertainty product in any Segal field and its
conjugate momentum. We have checked that the expectation value
of the metric (on Killing orbits) reproduces in fact the
classical solution that one would obtain from the expectation
value of the Segal field. However, the relative fluctuations in
this metric become huge for all coherent states when the
coordinate $u$ gets large and negative. Such values of $u$
describe the vicinity of a region where null cones are focused
by the plane wave \cite{PE}. Therefore, the quantum geometries
represented by coherent states do not admit an approximate
classical description in that region and classical spacetime
breaks down.

It is worth remarking that this result applies as well to the
vacuum of the model, which should correspond to flat spacetime.
In this sense, the spacetime foam around the vacuum turns out to
be quite significant, at least close to the focusing region.
This fact casts serious doubts on the possibility that, after
introducing matter fields in the system, a quantum field theory
around the flat-spacetime solution of our model could provide a
good semiclassical approximation \cite{WA} to the full quantum
theory.

It would be interesting to analyze the metric fluctuations for
families of quantum states other than coherent states.
Obviously, this analysis would not affect our conclusions about
the vacuum. In addition, note that, even if it were possible to
find states with smaller metric uncertainties, this would be
done at the cost of losing coherence in the Segal field.

On the other hand, the results attained might well depend on
several choices that have been made either in the construction
of the model or in the quantization process. Among such choices,
let us briefly comment on the representation selected for the
quantum theory, on the mode decomposition and regularization
adopted for the basic field $Y$, and on the system of
coordinates employed to describe the plane waves.

The commutation relations (\ref{acommuta}) admit representations
other than that discussed in Sec. V. For instance, one could
have made a different choice of creation and annihilation
operators that led to a unitarily equivalent Fock representation
but with a different vacuum. In this sense, the question that is
physically important is the correct identification of the state
represented by the vacuum in our quantum theory. This vacuum is
the only normalized state with minimum uncertainty in the Segal
field $\hat{Y}_R[f]$ and its canonical momentum (for all vectors
$f$) that is peaked around the zero field. Since spacetime is
flat when the classical field $Y$ vanishes, it seems reasonable
to identify our vacuum as the state corresponding to the flat
solution in the constructed quantum theory.

One can also find representations of our commutation relations
that are unitarily inequivalent to that employed in Sec. V. It
then could happen that our conclusions were not valid for some
of such representations. It is worth noticing, nevertheless,
that the Fock representation adopted guarantees that the Segal
field $\hat{Y}_R[f]$ (i.e., the basic field of the theory) is a
self-adjoint operator for all vectors $f$. In addition, the
existence of a cyclic \cite{RS} vacuum that describes the
classical flat solution obtained when the field $Y$ vanishes
makes of our choice of representation a natural selection. In
any case, the aim of this work is not at showing that a
classical description of the spacetime is precluded in quantum
gravity for all possible representations, but that the existence
of such an approximate description cannot be taken for granted
until the quantum theory is known.

On the other hand, in terms of the coordinate $u$, the expansion
in modes employed for our basic field $Y(u)$ is most natural,
since it is its standard Fourier expansion. Nevertheless, one
might have chosen to expand the field in the Fourier modes
associated with another coordinate $\bar{u}\in I\!\!\!\,R$,
related to $u$ by means of a (differentiable) bijective
transformation $u=H(\bar{u})$ [so that $H^{\prime}(\bar{u})\neq
0$]. One can always set $H^{\prime}(\bar{u})$ to be positive,
e.g., by using the invariance of the geometry under a reversal
of the null coordinates $V$ and $\bar{u}$. In addition, the
considered change of coordinate amounts in fact to a different
selection of function $z_0$ in Eq. (\ref{Phi}), namely,
$\bar{z}_0(u)\equiv z_0[H(u)]$. Remarkably, it turns out that
the gauge fixing and symmetry reduction carried out in Ref.
\cite{MM}, as well as the analysis and quantization performed in
the present work [except for definition (\ref{uU})], continue to
be valid with the replacement of $z_0(u)$ with $\bar{z}_0(u)$,
provided that the latter is a strictly increasing function that
ranges over the whole negative axis. One can easily see that the
function $\bar{z}_0(u)$ analyzed here satisfies these
conditions. As a consequence, adopting a Fourier expansion in
terms of the coordinate $\bar{u}$ can be considered equivalent
to the introduction of a different gauge fixing for the
$u$-coordinate diffeomorphisms, and leads just to an alternative
representation of our commutation relations that is carried by a
(possibly) different Fock space. Actually, it is not difficult
to show that such a representation is related to that introduced
in Sec. V by a Bogoliubov transformation of the creation and
annihilation operators. Moreover, one can check that all the
coherent states of the quantum theory that is obtained with this
alternative expansion in modes present also large relative
fluctuations in the metric when one approaches the focusing
region.

As for the regularization of our basic field, we recall that the
infrared divergences have been eliminated by imposing that the
field vanishes at $u_0$, both classical and quantum
mechanically, whereas the ultraviolet divergences have been
removed by averaging the field around the points $u$ and $u_0$
with a smooth function $g$ of compact support, an average that
amounts to a smearing of the field. The choice of the point
$u_0$ can be made on physical grounds: the value of the field at
that point is taken as a reference value for all measurements.
On the other hand, it seems reasonable to assume that the
support of the function $g$ is of the order of the Planck
length. The form of this function, however, remains quite
arbitrary. This introduces an ambiguity in the quantization
process, which is similar to that encountered in Ref. \cite{AP}
when defining the regularized metric for (the dimensional
reduction of) cylindrical waves with linear polarization. The
conclusion that the metric fluctuations become large for
coherent states when one approaches the region where null cones
are focused is nevertheless regularization independent, as we
argued in Sec. VI, and is basically due to the fact that the
metric on the two-dimensional Killing orbits degenerates in the
focusing region [so that $e^{z_0}$ vanishes in Eq.
(\ref{class})].

Our last comments refer to the system of coordinates employed.
This system does not cover the totality of the spacetime that
can be described with harmonic coordinates. Moreover, the region
where the metric uncertainty explodes ($u=-\infty$) corresponds
to a coordinate singularity in group coordinates, and this
singularity can be removed by introducing harmonic coordinates
\cite{VE}. One might then wonder whether the huge quantum
fluctuations detected could have been avoided by adopting
harmonic coordinates.

We remind, nonetheless, that the spacetimes described by those
coordinates are not globally hyperbolic \cite{PE}. Since the
canonical formulation of general relativity can only be applied
to spacetimes which can be foliated in time slices, the
quantization of gravitational plane waves using harmonic
coordinates cannot be achieved by standard methods. Our results
will be relevant for the whole of the spacetime covered with a
single chart in harmonic coordinates provided that one can
neglect the effect of the regions that are not described by our
system of coordinates. One expects this to be the case, at
least, for sandwich waves that leave the spacetime flat in those
regions. More importantly, one might have expected that the
quantization process would actually smooth out the strong
gravitational effects exerted by the plane wave in the region
where null cones are focused (i.e., in the vicinity of the
coordinate singularity in group coordinates), in a similar way
as it is usually expected that spacetime singularities disappear
in quantum gravity. On the contrary, and at least for the
spacetimes described by coherent states, our results prove that
such strong effects are in fact enlarged in the quantum theory.

\acknowledgments

The authors are grateful to P. F. Gonz\'{a}lez-D\'{\i}az for valuable
discussions and comments. They are also thankful to A. Roura and
E. Verdaguer for discussions. G. A. M. M. acknowledges DGESIC
for financial support under the Research Projects No. PB97-1218
and No. HP1988-0040. M. M. was supported by CICYT under the
Research Project No. AEN98-04031 and by funds provided by a
Basque Government FPI grant.


\begin{references}
\bibitem{note} We assume that the choice of  $u_0$ is such
that the distance $|u-u_0|$ is bounded from below by a strictly
positive constant for $u\ll 0$.
\bibitem{nota} Actually, the norm of
$f(u|u_0)$ generally diverges when $|u|\rightarrow\infty$. A
case in which the norm remains finite in that limit is when one
considers that what is kept fixed is the distance $|u-u_0|$,
rather than the location of the reference point $u_0$.
\end{references}
\end{document}